\documentclass[twocolumn,tighten,time]{aastex631}
\usepackage[fleqn]{amsmath}
\usepackage{graphicx}	
\pdfoutput=1 
\usepackage{epsfig}
\usepackage{epstopdf}
\usepackage{subfigure}
\usepackage{multirow}
\usepackage{hyperref}

\newcommand{\lav}{\langle}
\newcommand{\rav}{\rangle}
\newcommand{\modot}{M$_\odot$\ } 
\newcommand{\modotp}{M$_\odot$}
\def\HI{\hbox{H\hskip1.3pt$\scriptstyle\rm I\ $}}
\def\HIp{\hbox{H\hskip1.3pt$\scriptstyle\rm I$}}

\def\HIIp{\hbox{H\hskip1.3pt$\scriptstyle\rm II$}}

\def\HeIp{\hbox{He\hskip1.3pt$\scriptstyle\rm I$}}
\def\HeII{\hbox{He\hskip1.3pt$\scriptstyle{\rm II}\ $}}
\def\HeIIp{\hbox{He\hskip1.3pt$\scriptstyle{\rm II}$}}

\def\HeIIIp{\hbox{He\hskip1.3pt$\scriptstyle\rm III$}}
\def\Lyap{Ly$\alpha$} \def\Lya{Ly$\alpha$\ } \def\Lyas{_{\rm Ly\alpha}}
\def\G{_{\rm G}} \def\B{_{\rm B}} \def\D{_{\rm D}} \def\AGN{_{\rm AGN}} \def\rec{_{\rm rec}}
\def\escG{_{\rm esc,G}}
\def\escAGN{_{\rm esc,AGN}}
\def\der{{\rm d}} 

\begin{document}

\shorttitle{LAEs at Cosmic Dawn}
\shortauthors{Salvador-Sol\'e et al.}

\title{Lyman-$\alpha$ Emitting Galaxies (LAEs) at Cosmic Dawn: Implications and Predictions}

\author{Eduard Salvador-Sol\'e}\author{Alberto Manrique}
\affiliation{Departament de F\'\i sica Qu\`antica i Astrof\'\i sica, Institut de Ci\`encies del Cosmos. Universitat de Barcelona, E-08028 Barcelona, Spain}
\author{J. Miguel Mas-Hesse}
\affiliation{Centro de Astrobiolog\'\i a (CSIC-INTA), Departamento de Astrof\'\i sica, Madrid, Spain}
\author{Cristina Cabello}\author{Jes\'us Gallego}
\affiliation{Departamento de F\'\i sica de la Tierra y Astrof\'\i sica
\& Instituto de F\'\i sica de Part\'\i culas y del Cosmos, IPARCOS
Universidad Complutense de Madrid, 
Facultad CC F\'\i sicas, Ciudad Universitaria, E-28040 Madrid, Spain}
\author{Jos\'e Miguel Rodr\'\i guez-Espinosa}
\affiliation{Instituto de Astrof\'\i sica de Canarias, E-38205 La Laguna, and Instituto de Astrof\'\i sica de Andaluc\'\i a, E-18008, Granada, Spain.}
\author{Rafael Guzman}
\affiliation{Department of Astronomy, University of Florida
, 211 Bryant Space Science Center, PO Box 112055, Gainesville, FL 32611-2055, USA}

\email{e.salvador@ub.edu}



\begin{abstract}
The detection of Lyman-$\alpha$ emitting galaxies (LAEs) puts severe constraints on the reionization history. In this Paper we derive the properties of very high-$z$ LAEs predicted in the only two reionization scenarios shown in a previous Paper to be consistent with current data on 15 independent evolving global (or averaged) cosmic properties regarding luminous objects and the intergalactic medium and the optical depth to electron scattering of ionized hydrogen to CMB photons: one with a monotonic behavior, which is completed by $z=6$, as commonly considered, and another one with a non-monotonic behavior with two full ionization events at $z=6$ and $z=10$. We find that the \Lya luminosity functions of very high-$z$ LAEs are very distinct in those two scenarios. Thus, comparing these predictions to the observations that will soon be available thanks to new instruments such as the {\it James Webb Space Telescope}, it should be possible to unveil the right reionization scenario. In the meantime, we can compare the predicted redshift distribution and UV (or Lyman-$\alpha$) luminosities of very high-$z$ LAEs to those of the few objects already observed at $z>7.5$. By doing that we find that such data are in tension with the single reionization scenario, while they are fully compatible with the double reionization scenario. 
\end{abstract}



\section{Introduction}\label{intro}

The observation of very high-$z$ galaxies is a challenging endeavor with outstanding cosmological implications. That is particularly true for active star-forming Lyman-$\alpha$ (\Lyap) Emitters (LAEs) (e.g. \citealt{Oea08}). Were \Lya photons not absorbed by the neutral hydrogen present in the intergalactic medium (IGM) at those redshifts, LAEs could be observed out to the formation of the first galaxies. The way the observed abundance and other properties of LAEs vary with $z$ thus provides valuable information on the reionization history of the IGM \citep{Rea15,Boea15,Sea15,R21}. 

The fact that the equivalent width (EW) distribution of the \Lya emission line and escaping fractions of \Lya photons begin to decline at $z \sim 6$ indicates that reionization was completed by that redshift \citep{Huea10,Hea11,Kea11}. This conclusion is also consistent with the fact that the \Lya luminosity function (LF) remains roughly constant before $z=5.7$ \citep{Oea08}, and begins to decrease in an accelerated way until the maximum redshift ($z=7.3$) where it has been possible to estimate \citep{Shea12,Kea14}. Specifically, from $z=5.7$ to $z=6.6$ it decreases a factor $\sim 2$ \citep{Huea10,Kea14,Zea17}, and from $z=5.7$ to $z=7.3$ a factor $\sim 10$ \citep{Shea12,Kea14}. Such an evolution of the \Lya LF indicates that the mean neutral hydrogen fraction in the IGM $\lav x_{\rm HI}\rav$ is rapidly increasing after $z=6$, being $\sim 0.3$ at $z\sim 7$ \citep{Sc14,Kea18,Huea19}. Lastly, the detection success of LAEs among Lyman-break galaxies (LBGs) also declines from $z=6$ to $z=7.3$ \citep{Sc14,Bea15,Fea16} as expected from the increasing $\lav x_{\rm HI}\rav$. 

According to these trends, the LAE abundance would be naively expected to diminish since $z=6$ until vanishing a little beyond $z=7.3$. Yet, not only does the decline of the LAE detection success reverts by $z\sim 7.5$ \citep{Sc14,Kea14,Bea15,Fea16}, but there are nowadays about 20 spectroscopically confirmed LAEs up to very high-reshifts: four clustered objects at $z\sim 7.55$ \citep{Fea13,Jea20}, one pair at $z\sim7.74$ \citep{Tea20},
and isolated objects at $z=7.64$ \citep{Hea17}),
$z=7.66$ \citep{Soea16},
$z=7.68$ \citep{Vea22},
$z=7.77$ \citep{Jea20},
$z=7.88$ \citep{Jea20},
$z=7.94$ \citep{Jea20},
$z=8.38$ \citep{Lea17},
$z=8.67$ \citep{Lea22},
$z=8.68$ \citep{Zea15},
$z=8.78$ \citep{Lea21},
$z=9.11$ \citep{Hea18},
and even likely $z=9.28$ \citep{Lea21}.

The evolution of the LAE abundance at very high-$z$ must depend on the reionization process. Indeed, for LAEs to be visible at $z>6$ they must lie in large enough ionized cavities so that the emitted \Lya photons are cosmologically redshifted out of resonance before reaching the neutral IGM. Near $z=6$ ionized cavities are very large, and host many galaxies and active galactic nuclei (AGN). Clustered galaxies and AGN seem also to be responsible for the large ionized cavities hosting LAEs at $z\sim 7$ (e.g. \citealt{Rea20,Huea19}), and even up to $z\sim 7.75$ \citet{Jea20,Tea20}. However, there are neither very bright AGN nor substantial galaxy overdensities at $z\sim 8-9$. On the other hand, no other kind of ionizing sources, such as Population III (Pop III) stars, should do the job because, if they were already able to ionize very large cavities at those redshifts, reionization would be completed before $z=6$. Consequently, the most likely explanation is that such large ionized cavities are carved by isolated galaxies \citep{Loeb05}, with particularly high ionizing luminosities due to their low metallicity (\citealt{Fea16,JR16}), which would themselves be seen as LAEs. This scenario is consistent, indeed, with the little evolution observed at the bright end of the \Lya LF at $z\sim 7$ compared to the rapid decrease of the rest of the LF \citep{Sea16,Tea21}. In fact, the LFs at those redshifts seem to show an excess at the bright end respect to their best Schechter fit \citep{Meabis15,Zea17,Huea19}.

But the reionization history of the Universe is poorly known. As shown in \citet{Sea17}, hereafter SS+17, there are two very distinct solutions compatible with all currently available data on the global (or averaged) properties of luminous objects and IGM: one with a monotonic \HIp-reionization process completed at $z\sim 6$, as commonly considered, but also another non-monotonic one with two complete ionization events, a first one at $z\sim 10$, and a second and definitive one at $z\sim 6$. 

The aim of this Paper is to investigate whether the constraints imposed by the observed very high-$z$ LAEs can break that degeneracy. Using the AMIGA galaxy formation model employed in SS+17 to find those two reionization scenarios, we derive the properties of visible very high-$z$ LAEs in each of them, and show that they are indeed very different. We find that current data slightly favor double reionization, though a definite answer to that relevant issue must wait until more detailed observations, now feasible thanks to new facilities such as the {\it James Webb Space Telescope} (JWST), are gathered.

In section 2 we briefly describe the AMIGA model used in the present Paper to infer the intrinsic properties of LAEs in the two reionization scenarios found in SS+17. Those two scenarios are presented in section 3. The intrinsic properties of LAEs found in those scenarios are given in Section 4, and the correction of their \Lya luminosities for ISM- and IGM-absorption is carried out in section 5. The final \Lya luminosity functions (LFs) and other related properties of visible very high-$z$ LAEs to be compared to observations are provided in section 6. The discussion of these results and the conclusions of this study are given in section 7. Throughout the Paper we assume the $\Lambda$ Cold Dark Matter Universe with $\Omega_\Lambda = 0.684$, $\Omega_{\rm m} = 0.316$, $\Omega_{\rm b} = 0.049$, $h=.673$,
$n_{\rm s} = 0.965$, and $\sigma_8 = 0.831$ \citep{Planck16}.    

\section{The AMIGA model}

The {\it Analytic Model of Igm and GAlaxy evolution} (AMIGA) is a semi-analytic model (SAM) specifically designed to model the formation and evolution of luminous objects and their feedback on the IGM since the ``Dark Ages'' \citet{MSS15,Mea15}. AMIGA was used in SS+17 to constraint the reionization history of the Universe (see Sec.~3). In the present Paper, the intrinsic properties of LAEs in the two possible reionization scenarios found in SS+17 are used as input data. It is thus convenient to briefly explain that model.

Similarly to other SAMs and hydrodynamical simulations of galaxy formation, AMIGA monitors the mass growth of dark matter halos, the cooling of the gas they trap, its accretion onto the main central galaxy in each halo where it triggers star formation and feeds the growth of the AGN at its center, the interactions between that galaxy and the satellites accumulated into the halo via accretion and major mergers, and the feedback of all components into the IGM, by accounting for all relevant energy, mass, and metal exchanges between all the different components, and all heating-cooling, ionization-recombination, and molecular synthesis-dissociation mechanisms involved in the process. The main difference of AMIGA with respect to other SAMs and simulations is the novel strategy it uses in order to avoid or alleviate two main problems affecting all those methods described next, which makes it particularly well-suited to accurately study the formation and evolution of very high-$z$ objects.

The first problem arises from the fact that current SAMs and hydrodynamical simulations monitor the evolution of galaxies along many realizations of halo merger trees. (Note that there is no fundamental difference between both methods: SAMs use $N$-body simulations to built merger trees, and hydrodynamical simulations use analytic recipes similar to those included in SAMs to deal with the baryon physics at subgrid scale.) This procedure is very CPU-time and memory demanding, which forces SAMs and simulations to use a limited halo mass resolution (of $\ga 10^5$ \modotp) and to start calculations at a moderately high $z$ (of less than $z=10$) where luminous objects are already in place and have somewhat altered the IGM. That is, of course, an important drawback for the accurate study of {\it hierarchical galaxy formation at very high-$z$} since the first (Pop III) stars, with a crucial feedback onto the IGM, begin to form at a a much higher resdshift ($z\sim 30$) in tiny halos (with masses as small as $M \sim 10^3$ \modotp). Moreover, to save CPU-time, ``classical'' SAMs and simulations do not monitor the coupled evolution and luminous sources and IGM, but adopt the evolution of the IGM calculated independently by approximate analytic means. Thus, the modeling is not fully consistent. 

Modern SAMs (e.g. \citealt{Magg22}) and hydrodynamical simulations (e.g. the {\it First Billion Years} project, \citealt{Paar15}; the {\it Renaissance} simulation, \citealt{X16}; and {\it the Pop III Legacy} project \citealt{J19}) overcome some of these problems, but they still have limitations. Indeed, though they reach redhsifts higher than $z=30$, include molecular cooling and the formation of Pop III stars, and monitor self-consistently the entangled evolution of luminous objects and IGM, the halo mass resolution is still of order $10^5$ \modotp, and they deal with small volumes (of $\sim 5$ Mpc comoving side), which makes their results not fully representative of the whole Universe. Moreover, to save memory they do not monitor the evolution of the detailed components of normal galaxies. In fact, the main aim of these SAMs and simulations is to study specific effects of Pop III stars at cosmic dawn, rather than the properties of normal galaxies at very high-$z$ as needed here.

To avoid this problem AMIGA does not build many realizations of major mergers, but follows halo growth analytically by means of a powerful formalism called {\it the Confluent System of Peak Trajectories} (CUSP), which accurately recovers all halo properties found in simulations (see \citealt{SSM21} and references therein). It then interpolates the properties of halos and their baryonic content (galaxies and gas) in arrays of halo mass and redshift, which are progressively built from high- to low-$z$ and at each $z$, from low- to high-masses. The properties of halos in every node of the array are calculated by evolving them by accretion since their formation in a major merger of two progenitors (according to the theoretical progenitor and formation time distributions), whose properties are drawn by interpolation in the array previously built, i.e. with higher redshifts and lower halo masses. The process is started at an arbitrarily high redshift where halos have trivial conditions (they have just trapped more or less primordial gas depending on their mass), and at each $z$ at a low enough mass so that the corresponding halos have not trapped evolved gas yet. This procedure speeds up the calculations, which allows us to start the calculations from fully consistent initial and boundary conditions with arbitrarily high-mass resolution and small-time steps, and to accurately calculate the feedback of luminous objects in halos of all masses at every redshift.  

For the reasons that will be apparent in next Sections, it is important to note that Pop III stars form in neutral pristine regions, and ionize and pollute with metals their environments where normal galaxies subsequently form. Indeed, when Pop III stars explode via pair-instability supernovae, their halos lose most of their gas (except for very rare massive halos), so galaxies do not usually form on the top of Pop III star clusters (e.g. \citealt{WA07,WA08}), but they form ex-novo in halos traping gas in those ionized, metal-polluted environments, regardless of whether or not that previously harbored Pop III stars (these may also end up inside them by accretion), and develop AGN at their centers seeded by the black hole remnants of very massive Pop III stars (and afterwards accreted onto galaxies). This means that the properties and location of normal galaxies in those ionized metal-poluted regions are uncorrelated with those of Pop III stars, which keep on forming as long as there are neutral pristine regions in the IGM. That absence of correlation between Pop III stars and galaxies will play an important role in the properties of visible high-$z$ LAEs depending on the particular reionization history of the Universe.

The sponge-like (null genus) topology of the ionized IGM will also play an important role in our predictions (see Sec. 5). In AMIGA, the IGM is accurately treated as an inhomogeneous multiphase medium, including well-delimited neutral, singly ionized (\HIIp), and doubly ionized (\HeIIIp) regions, with different temperatures due to the action of luminous objects. When the ionized volume filling factor $Q_{\rm HII}=1-\lav x_{\rm HI}\rav$ is near 0.5, the IGM has a Swiss-cheese-like topology, with ionized cavities or neutral regions playing the role of holes (genus equal to $-1$) when $Q_{\rm HII}$ approaches zero or unity, respectively \citep{Lea08}. The singly/doubly ionized regions inside ionized cavities have a similar topology, depending on the doubly ionized volume filling factor $Q_{\rm HeIII}=1-\lav x_{\rm HeII}\rav-\lav x_{\rm HI}\rav$, though it is not important for our purposes here. As we will see below, the topology of ionized regions has important consequences in the properties of observable LAEs. 

\begin{table*}
\caption{Best fitting values of the free parameters$^*$ defining the two reionization models.}\label{bestparam}
\vspace{-15pt}
\begin{center}
\begin{tabular}{cccccccccccccc}
\hline \hline 
\smallskip
Model & $M_{\rm III}^{\rm lo}$ [\modot] & $\epsilon\B$ & $\epsilon\D$ & $\alpha\G$ & $\epsilon\AGN$ &
$h\rec$ & $f\escG^{\rm ion}$ & $f\escAGN^{\rm ion}$\\  
\hline
\smallskip
Single & $38^{+5}_{-7}$ & $1.00^{+0.00}_{-0.05}$ & $0.00^{+0.05}_{-0.00}$ & $0.33\pm 0.11$ & $0.0016\pm0.0003$ & $0.4\pm 0.3$ & $0.053\pm 0.007$ & $0.0053\pm 0.0006$ \\
\smallskip 
 Double & $87^{+9}_{-6}$ & $1.00^{+0.00}_{-0.05}$ & $0.00^{+0.05}_{-0.00}$ & $0.19\pm 0.04$ & $0.0025\pm 0.0004$ & $0.4\pm 0.3$ & $0.055\pm 0.008$ & $0.0053\pm 0.0005$ \\
\hline
\end{tabular}
\end{center}
$^*M_{\rm III}^{\rm lo}$: lower mass of Pop III stars with a top-heavy Salpeter-like IMF (the upper mass and the slope of the IMF are fixed through consistency arguments (see SS+17 for details); $\alpha\G$: the star formation efficiency; $\epsilon\B$ and $\epsilon\D$: the supernovae heating efficiencies in spheroids and disks, respectivel; $\epsilon\AGN$: the AGN heating efficiency; $f\escG$ and  $f\escAGN$: the escaping fractions of \HIp-ionizing photons from galaxies and AGN, respectively; and $h\rec$: the intra-halo mixing reheated gas mass fraction.
\smallskip\smallskip 
\end{table*}

\begin{table}
\caption{Free parameters$^*$ fixed by consistency arguments.}\label{otherparam}
\begin{center}
\vspace{-15pt}
\begin{tabular}{ccccc}
\hline \hline 
\smallskip
Model & $Z_{\rm c}$ & $M_{\rm III}^{\rm up}$ & $\alpha_{\rm III}$ & $\log(\rho_{\rm dis})$ \\
\smallskip
      &  [Z$_{\odot}$] &  [\modotp]            &                       &    [Mpc$^{-3}$]\\
\hline
\smallskip
Single & $-4.6\pm0.1$ & 
$300_{-0}^{+40}$ & $-2.35^{+0.10}_{-0.00}$ & $6.0\pm 0.6$ \\
Double & $-4.0\pm0.1$ & 
$300_{-0}^{+40}$ & $-2.35^{+0.10}_{-0.00}$ & $6.0\pm 0.6$ \\
\hline
\end{tabular}
\end{center}
$^*Z_{\rm c}$: metallicity threshold for atomic cooling; $M_{\rm III}^{\rm up}$: upper mass of Pop III stars; $\alpha_{\rm III}$: logarithmic slope of Pop III star IMF; $\rho_{\rm dis}$: characteristic gas mass density in dissipative contraction. 
\end{table}

The second problem of all galaxy formation SAMs and hydrodynamical simulations arises from the way they deal with all poorly known baryionic physics involved in galaxy formation (at subgrid scale, in case of simulations). This is achieved through simple physically motivated recipes that involve a large number of free parameters. These parameters are calibrated by fitting a few observations, but such a ``tuning" is dangerous for two reasons. First, the quantities that are parameterized may not take fixed values in the real Universe. However, letting them vary with $z$ or be multi-valuated would not be of much help because the poor knowledge of the mechanism does not allow to foresee any particular evolution with $z$ nor to accurately relate its multiple values with other quantities. More importantly, the fact that the model is able to provide a good fit to a few observations does not guarantee the validity of the underlying physics; it could just be due to the large number of free parameters used. In other words, it does not guarantee that the model also fit any other observation.

This problem is mitigated in AMIGA by causally connecting as many mechanisms as possible (this is the case, e.g., of dissipative contraction of gas-rich galaxies or of AGN growth), and taking into account as many consistency arguments as possible so as to have {\it the minimum possible number of independent free parameters to adjust}. At the same time, the calibration (tuning) of AMIGA is carried out by fitting {\it the maximum possible number of independent observations}. The price we must pay for this exigent procedure is the risk of finding no acceptable solution at all. But, if there is any, it will be more reliable than found in other less exacting models. 

\section{Single and Double Reionization}

In the version of AMIGA used in SS+17 to constrain the reionization history of the Universe, the number of free parameters reduced to 8 (see Table \ref{bestparam}). Their best values were adjusted through the fit to all currently available data on the evolution of global (or averaged) cosmic properties, namely the cold gas mass, stellar mass, and massive black hole (MBH) mass densities; the IGM, intra-halo gas, stellar, and inter-stellar metallicities; average galaxy sizes and morphological fractions; the star formation rate density; the galaxy and AGN ionizing emissivities; the IGM temperature in neutral and (singly and doubly) ionized regions; and galaxy and MBH mass functions, together with the optical depth to electron scattering of ionized hydrogen to CMB photons. The values of the remaining free parameters (see Table \ref{otherparam}) were fixed by consistency arguments (see SS+17 for details). In addition, we adopted a $z$-dependent clumping factor of the analytic form provided by \citet{Fea12}, though adapted to the actual redshifts of complete reionization and of formation of the first luminous objects; an IMF of ordinary Population I \& II stars with the Salpeter slope, $-2.35$, at large masses $M_\star$ up to 130 \modot, and the slope $-1$ for masses in the range $0.1$ \modot $< M_\star < 0.5$ \modot according to \citet{Wea08}; and 3) a radius-stellar mass relation for non-dissipatively contracted spheroids of the form: $r_{\rm B}=A M_{\rm B}^\gamma$, with $A \approx 0.049$ Kpc \modotp$^{-\gamma}$ and $\gamma\approx 0.20$, as suggested by observations \citet{Pea97,Sea07}. Even though there is, of course, some uncertainty in the preceding expressions, we checked the results to be very robust against reasonable variations. 

\begin{figure}
\centerline{\includegraphics[scale=0.53,bb= -290 110 872 440]{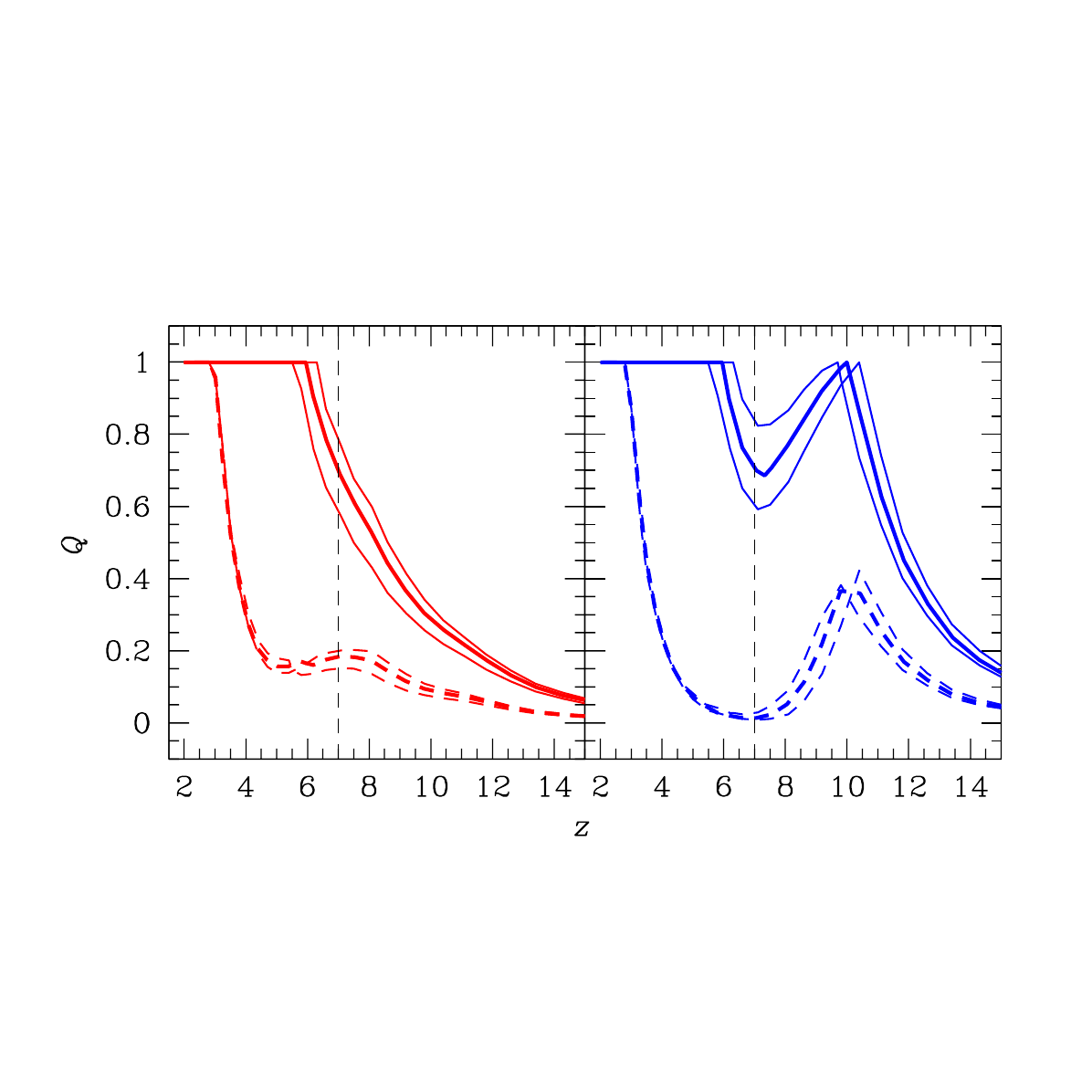}}
\caption{
\HI (solid lines) and \HeII (dashed lines) reionization histories expressed through their volume filling factors $Q_{\rm HII}=1-\lav x_{\rm HI}\rav$ and $Q_{\rm HeIII}=1-\lav x_{\rm HeII}\rav-\lav x_{\rm HI}\rav$ in single reionization (left panel) and double reionization (right panel) compatible with current global data on galaxies and AGN and the CMB radiation (SS+17). Thick lines give the best solutions of each kind, while thin lines bracket their acceptability range. The vertical dashed black lines mark $z=7$ where $Q_{\rm HII}$ is found to be $\sim 0.7$.}
\label{f1}
\end{figure}

The problem so posed was clearly overdetermined. Yet, the solution turned out to be degenerate. Two disjoint narrow sets of acceptable solutions were found giving similar good fits to the data: one with a monotonic \HIp-reionization completed at $z\sim 6$, and another non-monotonic \HIp-reionization with two complete ionization events, a first one at $z\sim 10$, and another definitive one at $z\sim 6$ (see Fig.~\ref{f1}). This is well-understood. \HIp-reionization is triggered by Pop III stars, but soon after the formation of the first generation of those stars normal galaxies begin to form. The contribution to ionization for the two kinds of objects depends on how top-heavy is the unknown Pop III star initial mass function (IMF). If it is very top-heavy, the contribution of normal galaxies (and AGN) to the ionizing emissivity is negligible in front of that of Pop III stars. Thus reionization is governed essentially by Pop III stars only. However, after full ionization, Pop III stars do not form anymore because the whole Universe has been not only ionized but also metal-polluted, so a recombination phase takes place until the increasingly abundant galaxies (and AGN) take over and lead to a second and definite full ionization of the IGM. For this second (or unique; see below) ionization to be completed by $z\approx 6$, the initial Pop III star-driven ionization must be completed at $z\approx 10$, which requires a specific quite top-heavy Pop III star IMF. Were this IMF top-heavier, the first ionization would be achieved at a higher $z$, and the optical depth to electron scattering of ionized hydrogen to CMB photons would exceed the empirical upper limit (SS+17 and references therein). On the contrary, were it less top-heavy, the first full ionization would be achieved at a lower $z$, and there would be no room for recombination to reach an ionized hydrogen fraction as low as $0.7$ at $z=7$ as observed \citep{Sc14,Kea18,Huea19}. Only in the case that the IMF were so little top-heavy that Pop III stars could not ionize the IGM by their own, there would be one only complete ionization at $z=6$ mainly driven by galaxies, with just a small contribution from Pop III stars (and AGN). In other words, the actual reionization scenario depends on the poorly known Pop III star lower-mass.

Similarly, the \HeIIp-reionization is initially driven by Pop III stars, while AGN take over at $z\sim 7$. The details of this process depend on the specific \HeIp-reionization history, but the result is qualitatively similar and satisfies the observational constraints in both cases.

In Fig.~\ref{f1} we see that the evolution of the ionized hydrogen fraction, $\lav x_{\rm HII}\rav$, below $z\sim 7.3$ essentially coincides in both reionization scenarios, but it strongly diverges at higher-$z$. While it keeps on decreasing in single reionization, it starts increasing again in double reionization after reaching a minimum at $z=7.35$. This means that the detection of very high-$z$ LAEs should thus be quite distinct in both scenarios. 

\section{Intrinsic LAEs}

For each reionization model, AMIGA supplies the evolving physical properties of IGM, but also those of luminous objects (Pop III stars, galaxies, and AGN) of all masses that lie in neutral, singly, and doubly ionized regions at any redshift between $z=60$ and $z=2$.\footnote{These are the predictions of the model in contrast with the evolving global (or averaged) cosmic properties used to tune it.} In particular, it provides the physical properties of star-forming galaxies (in ionized regions), as needed in the present study.

Specifically, the \Lya luminosities, $L\Lyas$, of these latter galaxies are calculated in AMIGA using the \citet{BC03} code that accounts for the nebular contribution to that line for any given metallicity, assuming a \citet{S55} IMF with upper mass limit of $M_\star=120$ \modotp.\footnote{The corresponding ionizing luminosity is simply given by the star formation rate for that metallicity. In the case of Pop III stars is instead calculated according to \citet{S02}.} We insist on that those \Lya luminosities are {\it intrinsic}, i.e. they do not account neither for the internal absorption by dust and the scattering by neutral hydrogen in the interstellar medium (ISM), nor for their external absorption by scattering by neutral hydrogen in the IGM. The correction for these absorptions must be carried out from some extra modeling, which will be done in section 5. This means that such ``intrinsic LAEs", defined for simplicity as all star-forming galaxies with an intrinsic $L\Lyas$ above $10^{38}$ erg s$^{-1}$, must not necessarily be visible. Next we describe the main properties of intrinsic LAEs predicted by AMIGA in the two alternate scenarios, used in the present study.

\begin{figure}
\centerline{\includegraphics[scale=0.44]{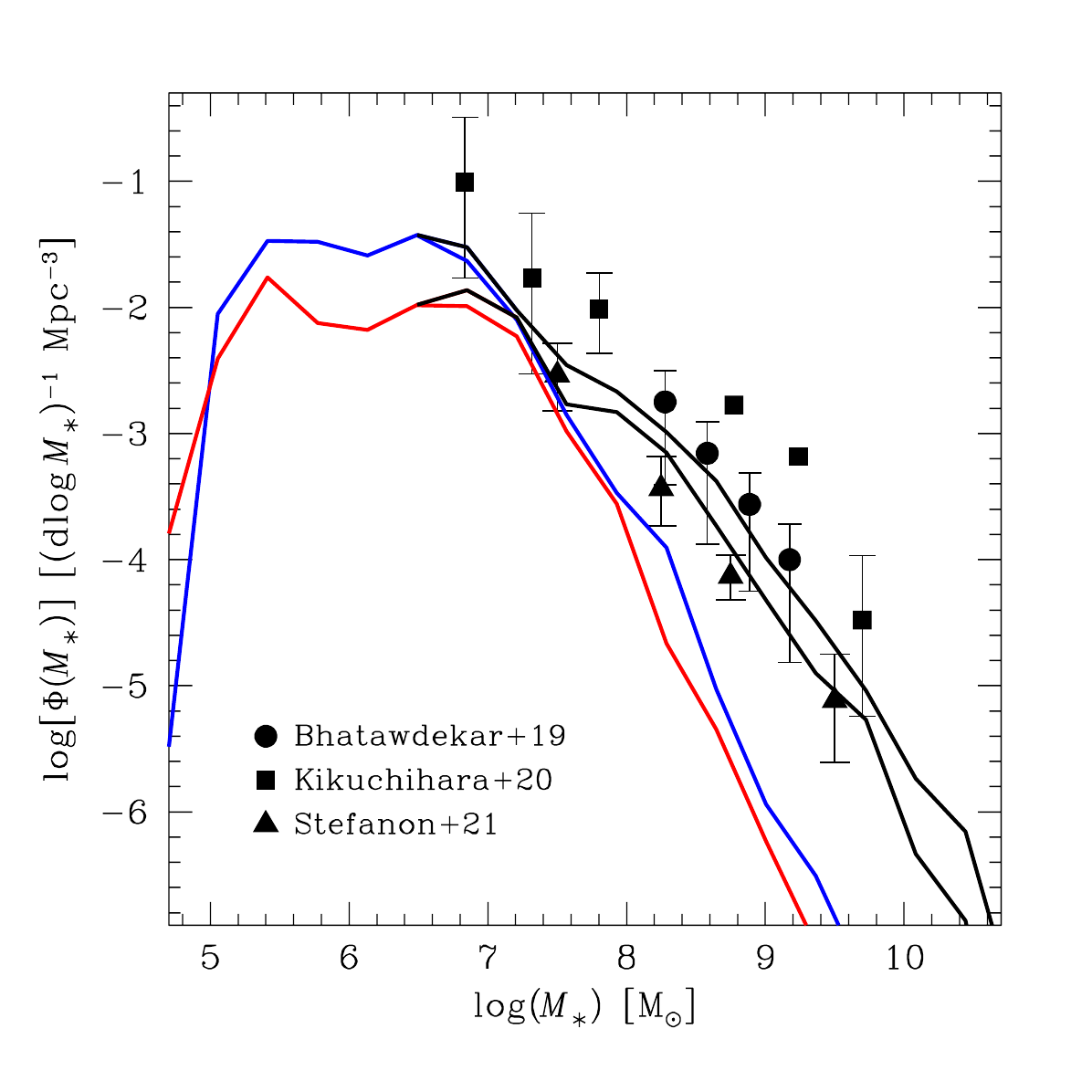}}
\caption{Differential stellar MFs of all galaxies (black lines) and LAEs (colored lines) predicted by AMIGA at $z=9$ in the two reionization scenarios. Specifically, the galaxy MF in single (double) reionization is the lower (upper) black curve and the corresponding LAE MF is the red (blue) curve. Black symbols give the observational estimates by different authors of the galaxy stellar MF drawn from the observed UV LF. (A color version of this Figure is available in the online journal.)}
\label{f2}
\end{figure}

\begin{figure}
\centerline{\includegraphics[scale=0.44]{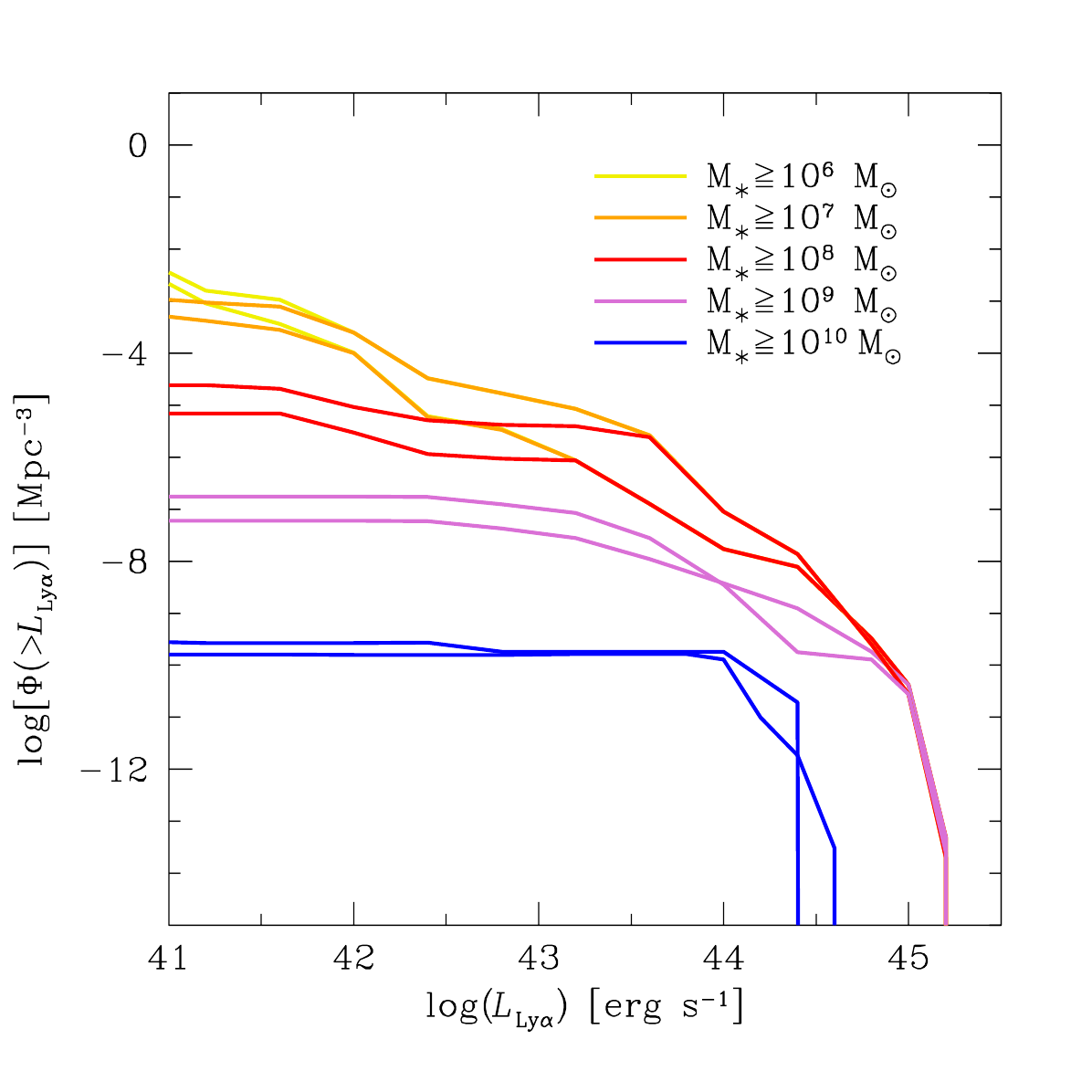}}
\caption{Contribution from LAEs of different stellar masses (colored lines) to the cumulative intrinsic \Lya LF predicted at $z=9$ in single and double reionization. The lower curves of each color correspond the most often to the case of single reionization, as expected from the slight difference between the global LAE abundances in the two reionization scenarios (see Fig.~\ref{f2}). (A color version of this Figure is available in the online journal.)}
\label{f3}
\end{figure}

\begin{figure*}
\includegraphics[scale=0.7,bb= -90 110 872 440]{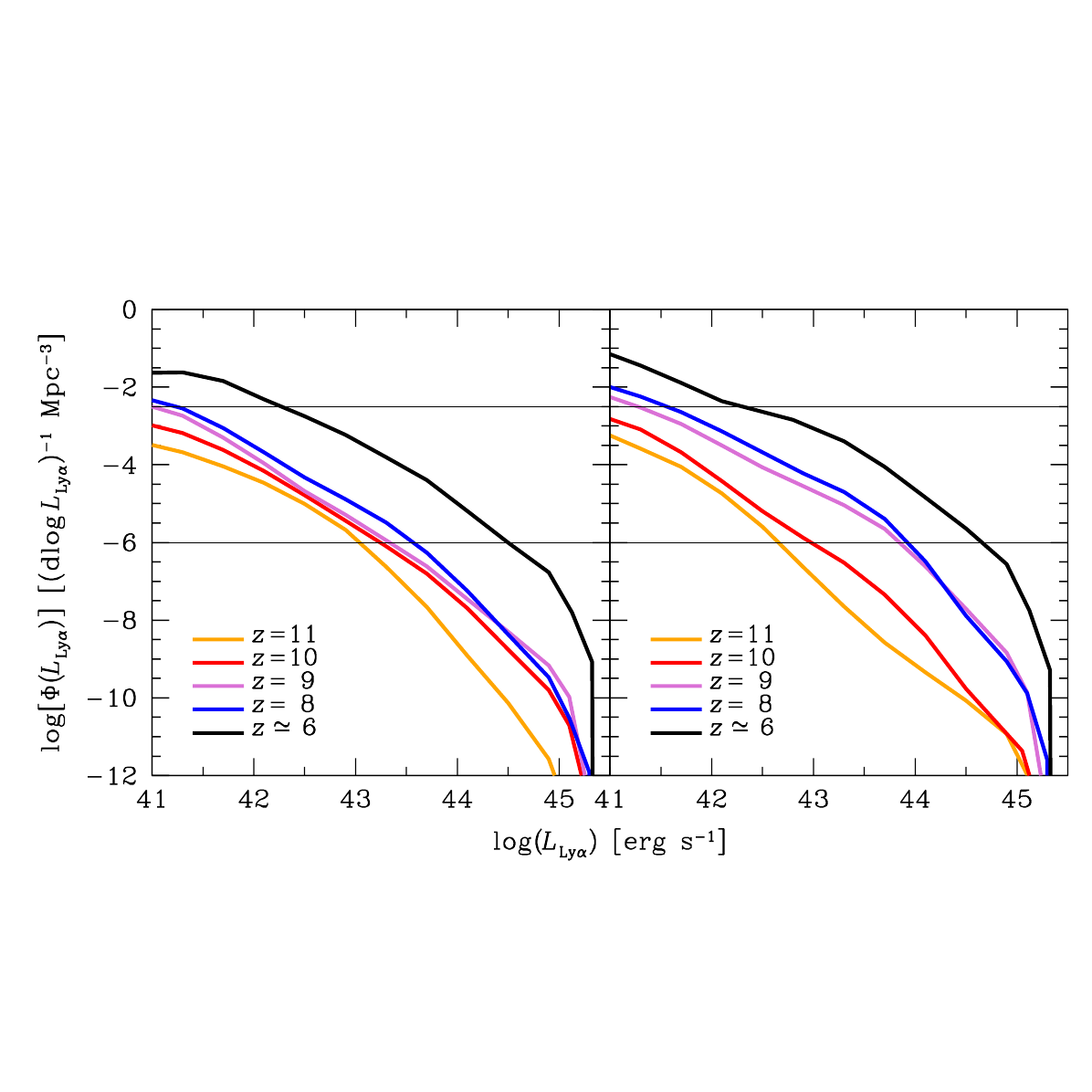}
\caption{Evolution of the differential intrinsic \Lya LFs (colored solid lines) in single (left panel) and double (right panel) reionization. At $z > 7.5$ the growth of LAEs of intermediate $L\Lyas$ in single reionization is somewhat delayed with respect to that in double reionization. The horizontal thin black lines bracket the ordinate range covered by current observations at high-$z$. (A color version of this Figure is available in the online journal.)}
\label{f4}
\end{figure*}

In Fig.~\ref{f2} we compare the (comoving) LAE stellar MFs and the global galaxy MF predicted at $z=9$, the highest redshift where the latter has been possible to measure from the galaxy UV LF \citep{Bea19,Kea20,Sea21}. The remarkable agreement between the predicted and observed galaxy stellar MFs gives much confidence to the predictions of AMIGA at very high-$z$. In addition, it illustrates that, as mentioned, the best models in the two reionization scenarios give similarly good fits to the observed global (or averaged) galaxy properties at all redshifts. In this Figure we see that the fraction of LAEs decreases smoothly with increasing mass, starting from unity at the low-mass end. The reason for this is simple: all low-mass galaxies are newborn objects which have formed from the rapid cooling of the gas recently trapped by low-mass (high-concentration) halos, while towards higher stellar masses there is a larger fraction of galaxies formed long time ago and having transformed most of their fuel in stars. 

The (comoving) LAE \Lya LF at $z=9$ is shown in Fig.~\ref{f3}. The brightest objects have a notable luminosity ($L\Lyas\sim 10^{45}$ erg s$^{-1}$), but their density is extremely low, so they will hardly be observed unless the selection function strongly favors high-$L\Lyas$ objects. The maximum $L\Lyas$ of LAEs of different stellar masses increases with increasing mass, reaches a maximum at $M_\star\sim 10^{9}$ \modotp, and diminishes again. That is well understood. The star formation rate is maximum in objects with the highest gas content and the smallest dynamical time. Since the latter corresponds to the maximum density reached by dissipative contraction, which is independent of galaxy stellar mass \citep{Mea15}, the maximum star formation rate of objects of a given stellar mass, $M_\star$, is simply proportional to $f_{\rm g}(M_\star)\,M_\star$, with $f_{\rm g}(M_\star)$ the highest gaseous to stellar mass ratio of galaxies of that mass. For the above mentioned reasons, $f_{\rm g}(M_\star)$ diminishes with increasing $M_\star$, so, even though more massive galaxies harbor more material, and hence, tend to form more stars, for high enough masses the decreasing gaseous to stellar mass ratio overcomes that trend. 

Intrinsic LAEs are similar in single and double reionization, but their evolution is somewhat different (see Fig.~\ref{f4}). Though the intrinsic \Lya LF are similar at both luminosity ends (in particular, the brightest objects are already in place at $z\sim 11$) and the LAE abundances show a similar slow increase with decreasing $z$ in parallel with that of galaxies in general in both scenarios, in double reionization there is in addition a puffing up at intermediate $L\Lyas$, which is absent in single reionization. More importantly, the growth of the LAE abundance is somewhat delayed in single reionization with respect to double reionization because, in the latter, ionized metal-polluted regions where galaxies form develop earlier. But, after full reionization at $z=10$, the increase of the LAE abundance in double reionization slows down as no new ionized metal-polluted regions are added, while this does not happen in single reionization until $z=6$. As a consequence, the LFs become very similar in both scenarios at $z\la 7$. 

\section{Visible LAEs}

\subsection{Correction for ISM-absorption}\label{ISM}

The modeling of internal absorption of the \Lya luminosity in LAEs is a complex subject that involves the structure and kinematics of neutral, molecular, and ionized hydrogen as well as dust around active star forming regions (see e.g. \citealt{Verea06}; \citealt{RTea15}; \citealt{Smea21}). It is thus out of scope correcting for the ISM-absorption by means of an accurate detailed model of that kind. Instead, we will apply a parameterized phenomenological correction fitting observations so that it should account in a statistical manner for all those internal effects. This is enough for our purposes here because, as mentioned, the different properties of LAEs in the two reionization scenarios we are looking for should arise from the distinct properties of ionized cavities found in each case rather than from the inner properties of LAEs themselves expected not to depend on the particular reionization history of the IGM.  

\begin{figure}
\centerline{\includegraphics[scale=0.44]{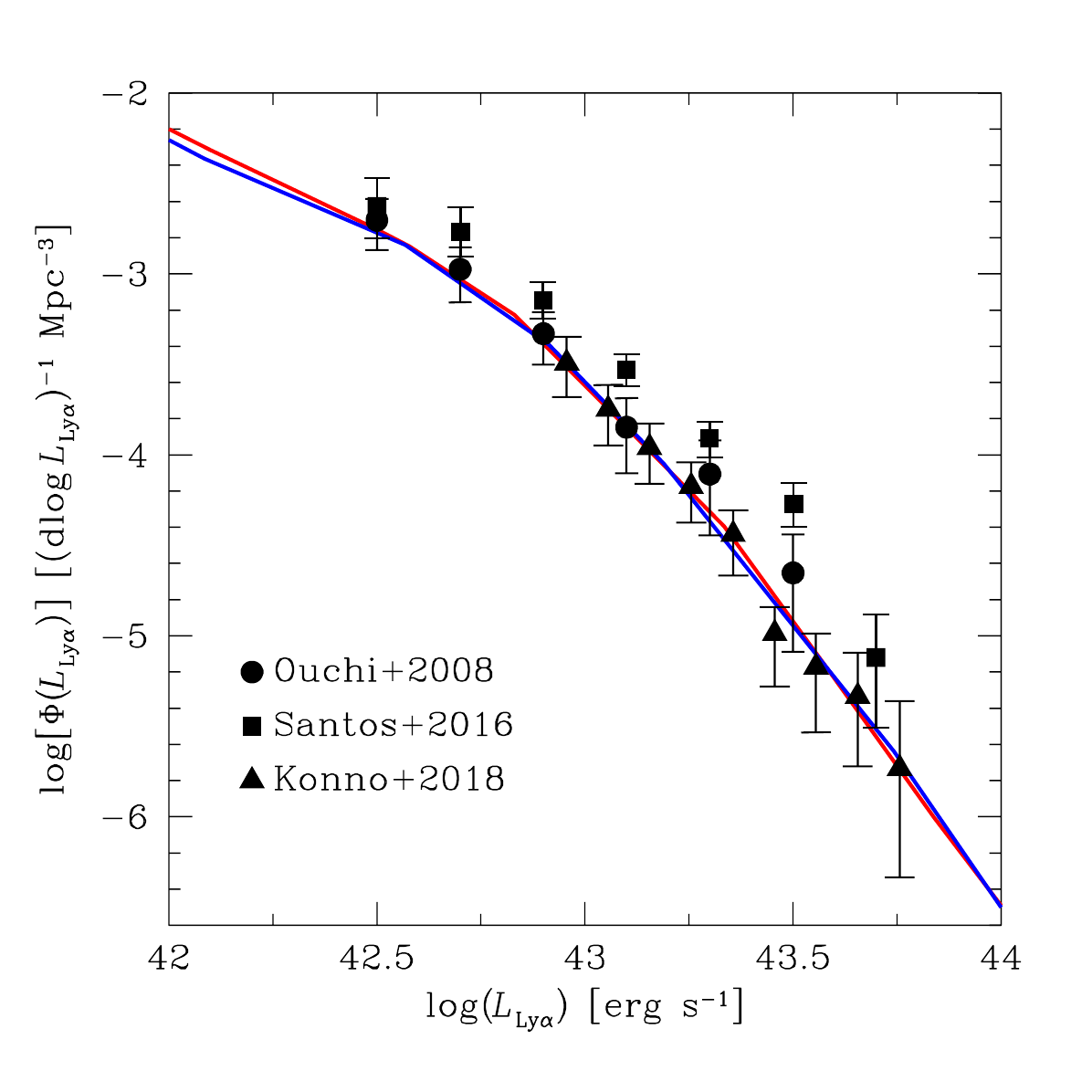}}
\caption{Differential \Lya LFs predicted at $z=5.7$ after correction for ISM-absorption in single (red lines) and double (blue lines) reionization compared to observations (dots with error bars). (A color version of this Figure is available in the online journal.)}
\label{f5}
\end{figure}

Due to internal absorption, the $L\Lyas$ of LAEs decreases a factor equal to the escaping fraction of \Lya photons, $f\escG^{{\rm Ly}\alpha}$. Were the escaping fraction the same for all $L\Lyas$, the corrected \Lya LF would be simply horizontally shifted by that fraction. Thus, by shifting the intrinsic \Lya LF at any given $z$ according to the global (averaged over all $L\Lyas$) $z$-dependent escaping fraction $f\escG^{{\rm Ly}\alpha}(z)\approx \min(1,\;5\times 10^{-4}(1+z)^{3.4})$ found by \citet{Hea11}, we should obtain a reasonable overall fit to the real LF. A small systematic deviation dependent on $L\Lyas$ should remain, however, due to the fact that, at each $z$, more massive galaxies formed earlier, so they have higher metallicities and more dust, in agreement with observation \citep{Dea16}. Thus the correcting factor we adopt is the above mentioned $z$-dependent global escaping fraction of \Lya photons, $f\escG^{{\rm Ly}\alpha}(z)$, times a simple power-law $(L\Lyas/L_0)^\nu$, with negative power index $\nu$. More precisely, to avoid that at very small $L\Lyas$ the LF so corrected may cross the intrinsic one, we take the minimum between both. As shown in Figure \ref{f5}, the values of $L_0$ and $\nu$ are equal to $10^{43.55}$ erg s$^{-1}$ and $-0.37$, respectively, in single reionization, and equal to $10^{43.15}$ erg s$^{-1}$ and $-0.31$, respectively, in double reionization, give excellent fits to the observed \Lya LF at $z=5.7$. This redshift is particularly well-suited for the fit because the \Lya LF is not affected by IGM-absorption, just by ISM-absorption as needed. 

Certainly, this correction involves the extrapolation beyond the observed ranges of the $z$-dependent global escaping fraction of \Lya photons, and, at each $z$, of the $L\Lyas$-dependent specific escaping fraction of \Lya photons. Thus, the real \Lya LFs could somewhat deviate from those derived here. However, the extrapolations used are of the less speculative form, i.e. simple power-laws, and lead to the expected value of unity of the escaping fractions of \Lya photons at $z\ga 9-10$ and, at each $z$, at the low-mass (low-$L\Lyas$) end, where galaxies have very low metallicities, and hence, very little dust. Thus the resulting \Lya LFs should be good approximations to the real ones. The most uncertain assumption on that corrections is that its dependence on $L\Lyas$ is the same at all redshifts. However, any possible deviation in that sense will only affect both luminosity ends of the LF at each $z$, which are unreachable to observation, so we can be unconcerned about. 

\subsection{Correction for IGM-absorption}

Strictly speaking, the shift produced in the wavelength of \Lya photons emitted by a LAE when they reach the surrounding neutral IGM depends not only on the separation $S$ between the galaxy and the edge of the ionized cavity as mentioned in the introduction, but also on other effects, such as the cosmological inflow onto the galactic halo, the peculiar velocity of the galaxy relative to the IGM, and the residual neutral fraction within the ionized cavity (see e.g. \citealt{GP04,Dea07,Mcea07,Lea11,MG20,Smea22}). However, while the former factor, with the most marked effect, is expected to yield a systematic difference at very high-$z$ in the two reionization scenarios due to the distinct correlation between the luminosity of visible LAEs and the size of ionized cavities carved in each scenario by different ionizing sources, all the remaining factors should have similar effects in both scenarios and, consequently, they are not expected to break the degeneracy between them. Moreover, as we will see below, even if they introduced a systematic effect added to that related to the separation $S$, it would not significantly affect the results. We will thus concentrate, hereafter, in the dependence of IGM-absorption on $S$ only, which greatly simplifies the treatment. 

The opacity of neutral hydrogen to the \Lya photons emitted by the LAE with wavelengths $\lambda_\alpha+\Delta \lambda$ ($\lambda_\alpha= 1215.67$ \r{A}) can then be approximated by the simple expression \citet{Loeb05}  
\begin{eqnarray}
\!\!\!&\tau&\!\!_\alpha\approx \frac{1.16\, {\rm pMpc}}{S}
\left(\frac{\Omega_{\rm b}/\Omega_{\rm m}}{0.17}\right) \label{1}\\ \nonumber 
&\times&\left[1\! -\! \frac{\Delta\lambda}{2.69\times 10^3 \textup{\r{A}}} \frac{10}{1+z}\log\!\left(\frac{2.44\times 10^4 \textup{\r{A}}}{\Delta\lambda}\frac{1+z}{10}\right)\!\right]\!,
\label{simple}
\end{eqnarray}
where $\Omega_{\rm m}$ and $\Omega_{\rm b}$ are the total matter and baryon density parameters, respectively. Thus, in the relevant wavelength range, the optical depth depends on $S$. Adopting the condition $\tau_\alpha=1$, equation (\ref{1}) leads to the minimum proper separation $S_{\min}(z)\approx 1.16$ pMpc for LAEs to be visible. Of course, this all-or-nothing condition is approximate: LAEs lying at a distance within the range $S_{\min}\pm \Delta S$ with $\Delta S\sim 0.6$ pMpc will be seen more or less dimmed (with $\tau_\alpha\sim 1\pm 0.5$). However, the volume $\Delta V$ occupied by such more or less dimmed LAEs is small compared to the volume occupied by all visible LAEs, i.e. by LAEs at a separation larger than $S_{\min}$, so we can neglect it, and adopt $S_{\min}$ as a clear-cut separation between LAEs contributing to the observable \Lya LF with essentially their intrinsic $L\Lyas$ corrected for ISM-absorption and absorbed ones. 

The way $S_{\min}$ enters the correction for IGM-absorption depends on whether or not the sizes of ionized cavities are correlated with the \Lya luminosity of visible LAEs, hereafter the $L\Lyas$-CS correlation. In the absence of correlation, the correction for IGM-absorption is simply achieved by multiplying the previous ISM-absorption-corrected \Lya LFs by the volume fraction occupied by visible LAEs, hereafter simply the visibility factor, $f_{\rm vis}$, equal to the volume fraction of ionized regions separated from the nearest foreground neutral region by more than $S_{\min}$. This correction is thus suited to double reionization, where cavities at very high-$z$ as needed here are ionized by massive Pop III stars uncorrelated with normal galaxies. (See below for redshifts $z\la 8.5$, when galaxies begin to reionize the IGM at the end of the recombination phase, and a small $L\Lyas$-CS correlation appears.)

Since in double reionization $\lav x_{\rm HI}\rav$ is substantially less than 0.5 at the redshifts of interest (see Fig.~\ref{f1}), the IGM must have a Swiss-cheese-like topology with small {\it neutral regions} embedded in an ionized background. Thus, $f_{\rm vis}(z)$ is simply one minus the volume fraction of neutral plus \Lya\!\!-shadowed regions,
\begin{eqnarray}
f_{\rm vis}(z)=
1-\int^1_0 \der s\,s^2\,[s+s_{\min}(z)]\,n(s,z)\nonumber\\
=1-\lav x_{\rm HI}\rav(z) -s_{\min}(z) \langle s^2 \rangle(z),~~
\label{new}
\end{eqnarray}
where $s$ is the size of neutral regions, $S$, scaled to the horizon diameter, $s=S/D_{\rm hor}(z)$, $s_{\min}(z)$ is $S_{\min}$ equally scaled, and $\lav s^2\rav(z)$ is the second order moment of $s$ for the probability density function of neutral bubble sizes $s$ at $z$, $n(s,z)$. To accurately calculate $f_{\rm vis}$ we would need the unknown function $n(s,z)$. However, we can still derive it with an error of less than 5\% for $\lav x_{\rm HI}\rav$ below 0.35 by taking into account that $n(s,z)$ is independent of the reionization history \citep{Lea08}. It must thus depend on $z$ through $\lav x_{\rm HI}\rav(z)$, so we can approximate $\lav s^2\rav$ in equation (\ref{new}) by its Taylor expansion to second order around full ionization ($\lav s^2\rav=0$): $\lav s^2\rav(z)=A\lav x_{\rm HI}\rav(z) +B\lav x_{\rm HI}\rav^2(z)$, with the values of constants $A$ ($A>0$) and $B$ such that $f_{\rm vis}$ satisfies the constraints given by the observed ratios of the $\Phi_*$ values in the Schechter fits to the \Lya LFs of visible LAEs at $z=6.6$ and $7.3$ with respect to that at $z=5.7$ \citep{Kea14}, undoing (for an asymptotic slope of $-2.0$) the horizontal shifts due to ISM-absorption to recover the meaning of the theoretical $f_{\rm vis}$ values. Certainly, two data points is not much. But we cannot do better because all observed \Lya LFs at $z> 6$ refer to essentially the same couple of redshifts ($z\sim 6.5$ and $z\sim 7-7.3$). Fortunately, we must determine only the values of two coefficients ($A$ and $B$), so two observational constraints are enough. We could only try to better determine these two data points using more estimates of $\Phi_*$. But the Schechter fits performed by all authors are unconstrained, i.e. the $\alpha$ and $L_*$ values are not fixed, and the $\Phi_*$ values so obtained cannot be used to constrain $f_{\rm vis}$. Only those provided by \citet{Kea14} were inferred keeping $\alpha$ and $L_*$ fixed as needed.\footnote{As mentioned, the correction for IGM-absorption when there is essentially no $L\Lyas$-CS correlation as at $z\la 7.3$, is carried out by multiplying by $f_{\rm vis}$ the \Lya LF corrected for ISM-absorption, i.e. with the same values of $alpha$ and $L_*$ as the intrinsic LF.} On the other hand, the LFs found by different authors at those redshifts are very similar, so the average $\Phi_*$ values we could obtain from the constrained fits to the raw data provided by those authors would be very similar to those found by \citep{Kea14}. 

\begin{figure}
\centerline{\includegraphics[scale=0.53,bb= 10 120 572 440]{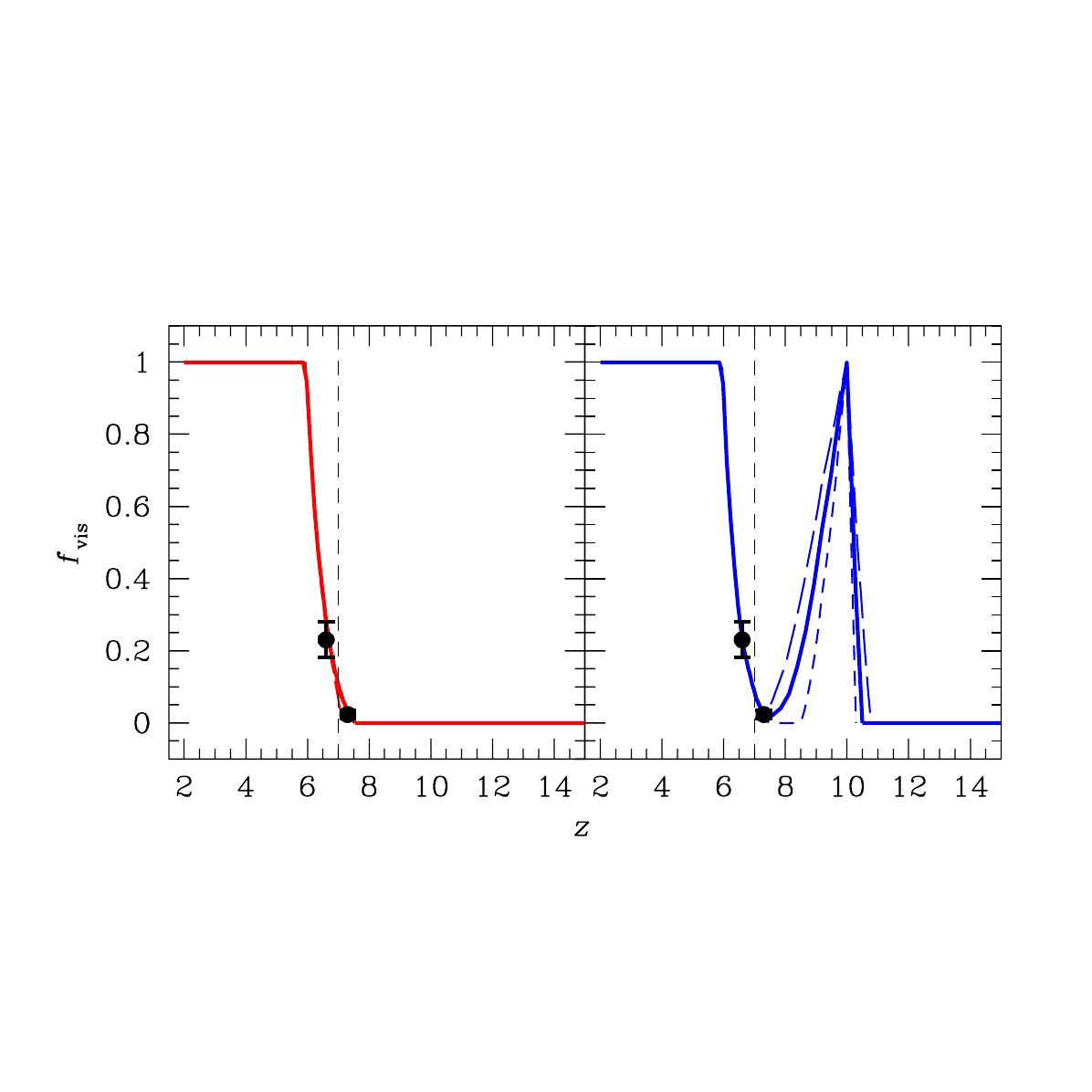}}
\caption{Visibility factor giving the best fit to the empirical estimates of $f_{\rm vis}$ found by \citet{Kea14} (dots with error bars) in single (left) and double (right) reionizations (thick solid lines). Thin short-dashed and long-dashed lines give the solutions we would obtain if, instead of taking $s_{\rm min}(z)\propto 1+z$ as found for $S_{\rm min}=1.16$ pMpc, we took take $s_{\rm min}(z)\propto (1+z)^2$ and $s_{\rm min}(z)$ constant, respectively, to mimic what could result from the dependence on other factors of the transmission of \Lya photons in neutral IGM. (At $z\la 7.5$ the dashed lines almost overlap with the solid ones in both reionization scenarios).}
\label{f6}
\end{figure}

In the right panel of Fig.~\ref{f6} we show the function $f_{\rm vis}$ so obtained (for $A=4650$ and $B=-7130$). As expected, it is unity at the two full ionization events (at $z=6$ and $z=10$), and has a minimum at the maximum of $\lav x_{\rm HII}\rav$. Lastly, after $z=10$ it decreases monotonically until vanishing by $z\sim 10.25$ ($\lav x_{\rm HI}\rav\sim 0.18$). This is thus the maximum redshift where LAEs can be detected in the double reionization scenario. Strictly speaking, the real maximum redshift of visible LAEs is expected to be somewhat higher because, when $f_{\rm vis}$ approaches zero, equation (\ref{new}) slightly underestimates it due to the overcrowding of the $L\Lyas$-shadowed zones behind neutral regions causing them to slightly overlap.\footnote{This does not mean, of course, that there can be no star-forming LBGs at substantially higher redshifts (see e.g. \citealt{Pea22,Hea22}), but simply that they cannot be seen as LAEs.} But, apart from this small flaw near $f_{\rm vis}=0$, the visibility factor is very robust since it relies on the reionization history (compare Figs.~\ref{f6} and \ref{f1}), through simple geometrical arguments with no need to model ionized cavities. 

As mentioned, in the case of a significant $L\Lyas$-CS correlation, as found in single reionization where the ionizing sources are galaxies themselves, $f_{\rm vis}$ cannot be used to correct the \Lya LF for IGM-absorption. However, it still measures the visibility zone of LAEs because the way it is calculated does not depend on the presence or not of that correlation or, equivalently, on the nature of ionizing sources. $f_{\rm vis}$ can thus be calculated in single reionization as well. The result is shown in the left panel of Fig.~\ref{f6}. At $z< 7.3$, $f_{\rm vis}$ is almost identical to that found in double reionization (right panel). But, at higher $z$, instead of having a minimum and then increasing, it keeps on decreasing until vanishing at $z\sim 7.5$ or so. (For the same reason above, the actual redshift where $f_{\rm vis}$ vanishes is expected to be somewhat higher, though the ever monotonic decreasing trend with increasing $\lav x_{\rm HI}\rav$ (or increasing $z$) is kept, implying that LAEs should rapidly disappear not much farther than $z\sim 7.5$.) This makes a great difference from the double reionization case.

The previous functions $f_{\rm vis}(z)$ have been obtained under the simplifying assumption that the transmission of \Lya photons in the IGM depends on $S$ only. We may thus wonder if they would change had we included the other factors entering that transmission mentioned above. The answer is that $f_{\rm vis}(z)$ is very insensitive to all these factors. They only affect $s_{\rm min}(z)$, while the solution of equation (\ref{new}) is quite insensitive, indeed, to that quantity. If $s_{\rm min}(z)$ were varied by any arbitrary constant factor, the change would be absorbed by the new values of coefficients A and B (see eq.~[\ref{new}]), so we would be led to exactly the same solution $f_{\rm vis}(z)$. While any (reasonable) variation in the dependence of $s_{\rm min}(z)$ on $z$ would yield a non-null though very small effect on the solution $f_{\rm vis}(z)$. This is shown in Fig.~\ref{f6}, where we depict the solutions resulting from changing the dependence $s_{\rm min}(z)\propto 1+z$, as found in case of a fixed separation of $S_{\rm min}$ equal to 1.16 pMpc, to $s_{\rm min}(z)\propto (1+z)^{1\pm 1}$. As can be seen, the new solutions $f_{\rm vis}(z)$ are very similar, indeed, to the original one. The reason for this is that $f_{\rm vis}(z)$ is almost fully determined by the reionization history ($\lav x_{\rm HI}\rav(z)$), with only a very small influence of $s_{\rm min}(z)$.

Let us turn now to the correction for IGM-absorption in single reionization, where ionized cavities are mainly carved by galaxies themselves. Due to the increasing $L\Lyas$-CS correlation at $z>7.3$,\footnote{Below that redshift ionized cavities are so large that they host many LAEs of all luminosities, and there is almost no correlation.} the decrease in the LAE abundance will be accompanied by their increasing brightening. Indeed, the IGM density increases and galaxies must be increasingly luminous to ionize large enough cavities. In this case, the correction of the \Lya LF for IGM-absorption in single reionization must account for that trend. Note that the same effect though much weaker is expected in double reionization at $z \la 8.5$, when recombination in the cavities previously ionized by Pop III is stopped by the ionization driven by galaxies (and AGN). We will come back to that particular case below. Here we concentrate in the case of single reionization.

When $\lav x_{\rm HI}\rav$ becomes substantially greater than 0.5 (at $z\ga 9$; see Fig.~\ref{f1}), there must only be ionized bubbles around isolated galaxies. But before that, at $z\ga 7.75$, the same condition already holds for ionized cavities around {\it visible LAEs}. Indeed, at those redshifts $\lav x_{\rm HI}\rav$ is still smaller than $0.5$, so ionized cavities still form a web of interconnected filaments. But these filaments are thin since basically populated by UV faint galaxies, so they do not harbor visible LAEs. Only a few nodes around UV bright galaxies are thick enough for their central galaxy to be seen as a LAE. And at $z\sim 7.75$ those UV bright galaxies are expected to be quite isolated.\footnote{Below $z=7.75$ LAE pairs are still frequent \citep{Jea20,Tea20}.} Thus, the proper radii $R$ of those ionized nodes must evolve with time $t$ according to the differential equation for {\it smooth} ionization (i.e. with no percolation with similarly large ionized nodes) around essentially isolated galaxies,
\begin{equation}
\frac{\der \left(\!R/a\!\right)^3}{\der t}=\frac{3f\escG^{\rm ion}L_{\rm ion}}{4\pi n_{\rm HI}(t)}-\frac{R^3(t)\alpha_{\rm HII}(t)\,C(t)\,n_{\rm HI}(t)} {a^6(t)},
\label{inieq}
\end{equation}
where $L_{\rm ion}$ is the (intrinsic) ionizing luminosity, in photons $s^{-1}$), of the central UV bright galaxy, $f\escG^{ion}$ is the escaping fraction of ionizing photons, $a=1/(1+z)$ is the cosmic scale factor, $n_{\rm HI}$ is the comoving mean neutral hydrogen density, $\alpha_{\rm HII}$ is the temperature-dependent recombination coefficient to \HIp\ (for the mean cosmic temperature in ionized regions at that redshift), and $C$ is the clumping factor. 

\begin{figure}
\centerline{\includegraphics[scale=0.44]{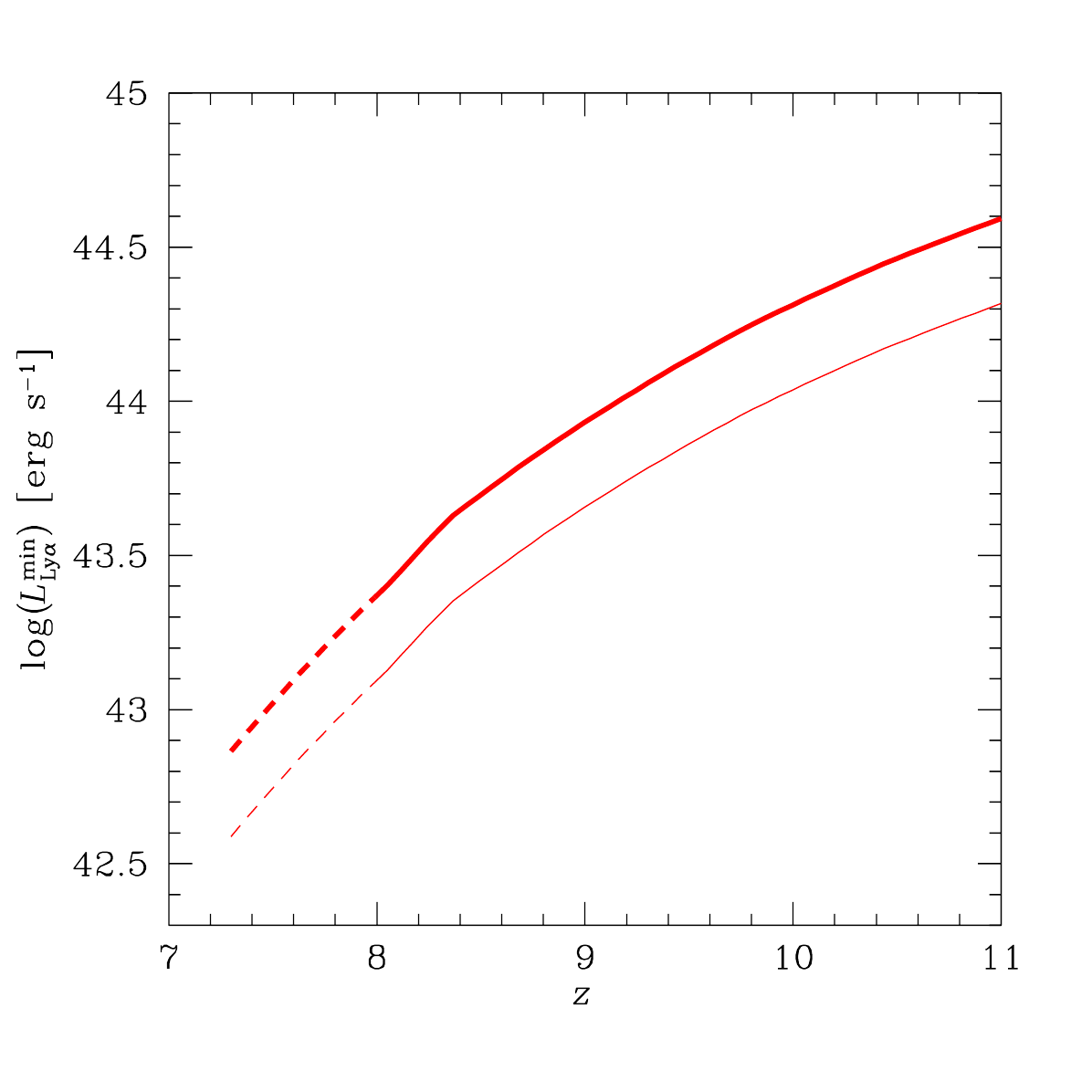}}
\caption{Minimum \Lya luminosity of LAEs in single reionization at very high-$z$ where ionized bubbles are carved by individual UV bright galaxies. We show the results obtained from the best value of the escaping fraction of ionizing photons, $f\escG^{\rm ion}=0.054$, found in SS+17 (thick line), and a more conservative value of $f\escG^{\rm ion}=0.1$ (thin line).}
\label{f7}
\end{figure}

The most favorable case for a UV bright galaxy to be seen as a LAE is when the ionized cavity (hereafter, the ionized bubble) around it is large, rather than when its ionizing luminosity $L_{\rm ion}$ is high. That distinction is important because, in a violent short-lived star formation burst, the radius of the ionized bubble is rapidly increasing, but it still is much smaller than if that $L_{\rm ion}$ had been operating for a long time. On the contrary, $R$ is maximum when $L_{\rm ion}$ has been kept constant for a long enough time for the comoving volume of the ionized bubble to reach quasi-equilibrium. Consequently, for a galaxy with $L_{\rm ion}$ to have chances to be seen as a LAE it must satisfy equation (\ref{inieq}) with null time derivative. Thus, taking $R$ equal to $S_{\min}$ in equation (\ref{inieq}) with $\der (R/a)/\der t=0$ we obtain the typical minimum ionizing luminosity of visible LAEs at $z$,
\begin{equation}
L_{\rm ion}^{\min}=\frac{4\pi}{3} S_{\min}^3(z)\!\left[\frac{X\rho_0\Omega_{\rm b}\lav x_{\rm HI}\rav(z)}{m_{\rm p}(1+z)^{-3}}\right]^2\!\frac{\alpha_{\rm HII}(z)C(z)}{f\escG^{\rm ion}},
\label{lion}
\end{equation}    
where $\rho_0$ is the (comoving) mean cosmic density, $m_{\rm p}$ is the proton mass, and $X$ is the primordial hydrogen mass fraction. Apart from $S_{\min}(z)$ given above, all the remaining quantities in equation (\ref{lion}), available electronically from \url{https://cdsarc.unistra.fr/ftp/J/ApJ/834/49} were obtained in SS+17 (see their Fig.~7) as some of the model predictions compared to observations.

Once $L_{\rm ion}^{\min}$ is known we can readily infer (for an EW of the intrinsic \Lya line of 100 \r{A} as found for LAEs in steady state since at least $\sim 20$ Myr; e.g. \citealt{BCD21}) the minimum intrinsic \Lya luminosity of the corresponding LAEs, and, applying the correction for ISM-absorption described above, determine the desired minimum \Lya luminosity, $L\Lyas^{\min}(z)$, of visible LAEs in single reionization at $z\ga 7.75$. (A similar derivation leads to the minimum UV luminosity, $L_{UV}^{\min}(z)$, of visible LAEs used in section 6.) 

The evolution of $L\Lyas^{\min}(z)$ from the best value of $f\escG^{\rm ion}$ equal to $0.054$ found in SS+17\footnote{We adopt this value valid for both single and double reionization; see Table 1.} is depicted in Fig.~\ref{f7}. That value of $f\escG^{\rm ion}$ is below the lower limit of $\sim 0.2$ found in theoretical studies (e.g. \citealt{Rea15,Boea15}) by enforcing that the ionization of the IGM at $z=6$ is carried out by galaxies alone. But AMIGA takes into account that Pop III stars and AGN also contribute to the ionization of the IGM, which explains that $f\escG^{\rm ion}$ is somewhat lower. In fact, the value of 0.054 fully agrees with all direct observational estimates (spanning from 0.03 to 0.07) found at $z\la 5$ (e.g. \citealt{Wea10} and references therein). None the less, since $f\escG^{{\rm Ly\alpha}}$ increases with increasing $z$ (see section \ref{ISM}), $f\escG^{\rm ion}$ could also, so we also show in Fig.~\ref{f7} the result we would obtain from the more conservative value of $f\escG^{\rm ion}=0.1$.

Therefore, the correction for IGM-absorption of the \Lya LF of LAEs (corrected for ISM-absorption) in single reionization at any $z\ga 7.75$ must be carried out by truncating it at the corresponding minimum luminosity, $L\Lyas^{\min}(z)$. Notice that, integrating the volume fraction in ionized nodes around visible LAEs at each $z$, we could derive a more accurate estimate of $f_{\rm vis}$ when it approaches zero than that given above affected by the overcrouding of \Lya-shadowed regions behing neutral ones.

The previous correction for IGM-absorption in the presence of $L\Lyas$-CS correlation accounting for large ionized nodes around LAEs also has repercussions on double reionization at $7.75 \la z\la 8.5$. Since UV bright galaxies with $L_{\rm ion}$ above $L_{\rm ion}^{\min}$ can then carve by themselves large enough ionized cavities for them to be visible as LAEs, they must be visible regardless of whether or not their ionized cavities were previously ionized (at $z> 10$) by Pop III. In other words, if the \Lya LF after correction for ISM-absorption in double reionization is higher at some $L\Lyas$ than its counterpart in single reionization, it cannot go below it after correction for IGM-absorption. If this happens when multiplying the \Lya LF by $f_{\rm vis}$, we must increase such a corrected \Lya LF until the inconsistency disappears. 

\section{Results}

\begin{figure*}
\includegraphics[scale=0.65,bb= -90 110 872 440]{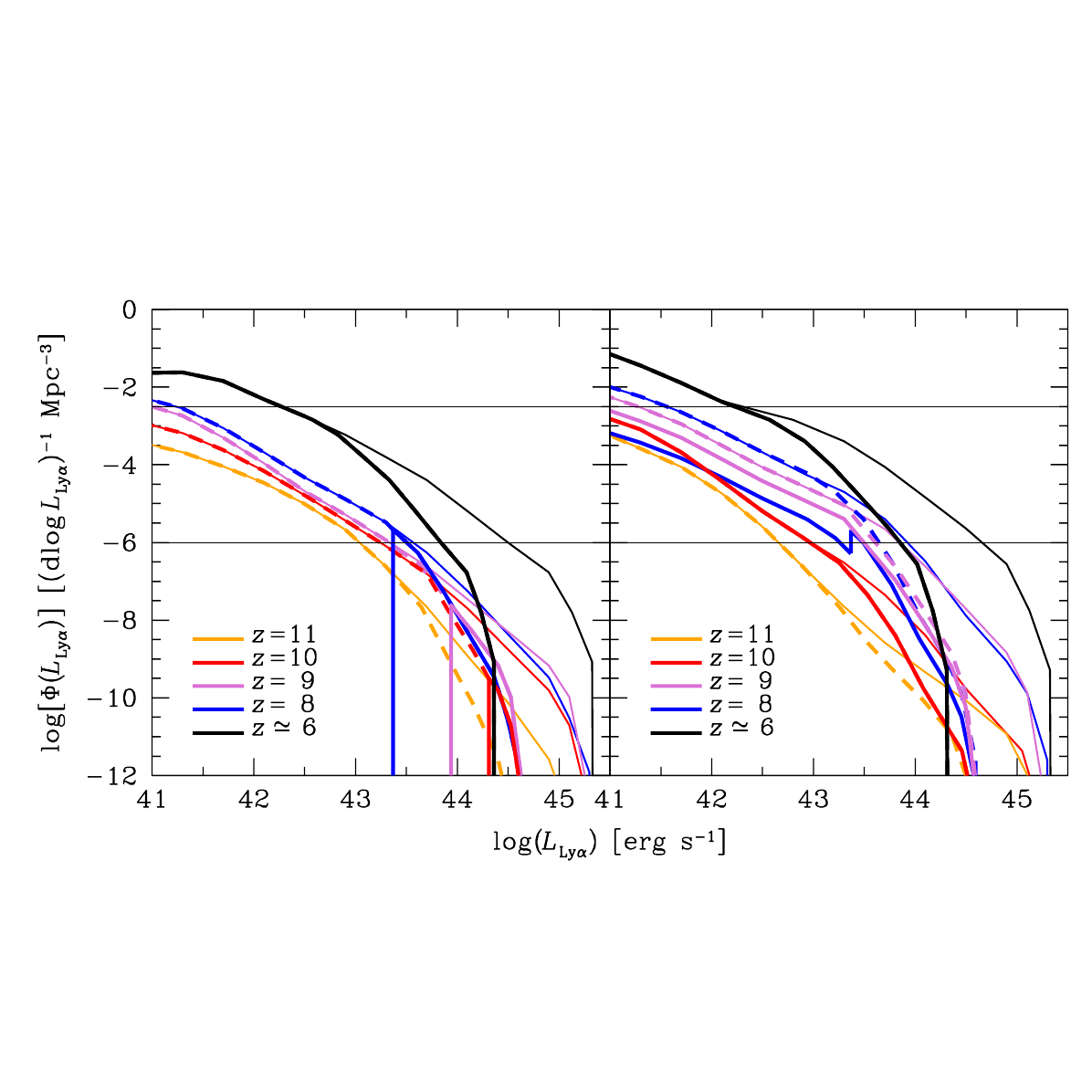}
\caption{Same as Fig.~\ref{f4} but for the \Lya luminosities corrected for ISM- plus IGM-absorption (thick solid lines) in single reionization (left panel) and double reionization (right panel). For comparison we show the original intrinsic \Lya LFs (thin solid lines) and after correction for ISM-absorption only (thick dashed lines). Note that there is no thick solid orange line. (A color version of this Figure is available in the online journal.)}
\label{f8}
\end{figure*}

The \Lya LFs of visible LAEs predicted in the two reionization scenarios are plotted in Fig.~\ref{f8}. (The small peak at log($L\Lyas\sim 43.4)$ in double reionization in the LF at $z=8$ is caused by the above mentioned refined correction for IGM-absorption.) As can be seen, after correction for IGM-absorption the LF at $z=11$ disappears in both reionization scenarios though for a different reason in each case. But this is the only coincidence in both reionization scenarios. In all the remaining redshifts the respective LFs greatly differ from each other. In double reionization, the increasing visibility of LAEs when $z$ approaches 10 balances the decreasing abundance of galaxies, causing the abundance of visible LAEs of any given $L\Lyas$ to stay rather constant in that redshift interval. On the contrary, the LFs in single reionization are truncated at a progressively higher $L\Lyas$ towards high redshifts. Thus, the comparison of the predicted \Lya LFs with the real ones at very high-redshifts, which will soon be possible to determine thanks to the new observational facilities, is a clear-cut test for unraveling the reionization scenario. In the meantime, however, we must be satisfied with the partial information brought by the redshifts and luminosities of the few very high-$z$ LAEs currently detected.

\begin{figure}
\centerline{\includegraphics[scale=0.50,bb= 20 140 572 470]{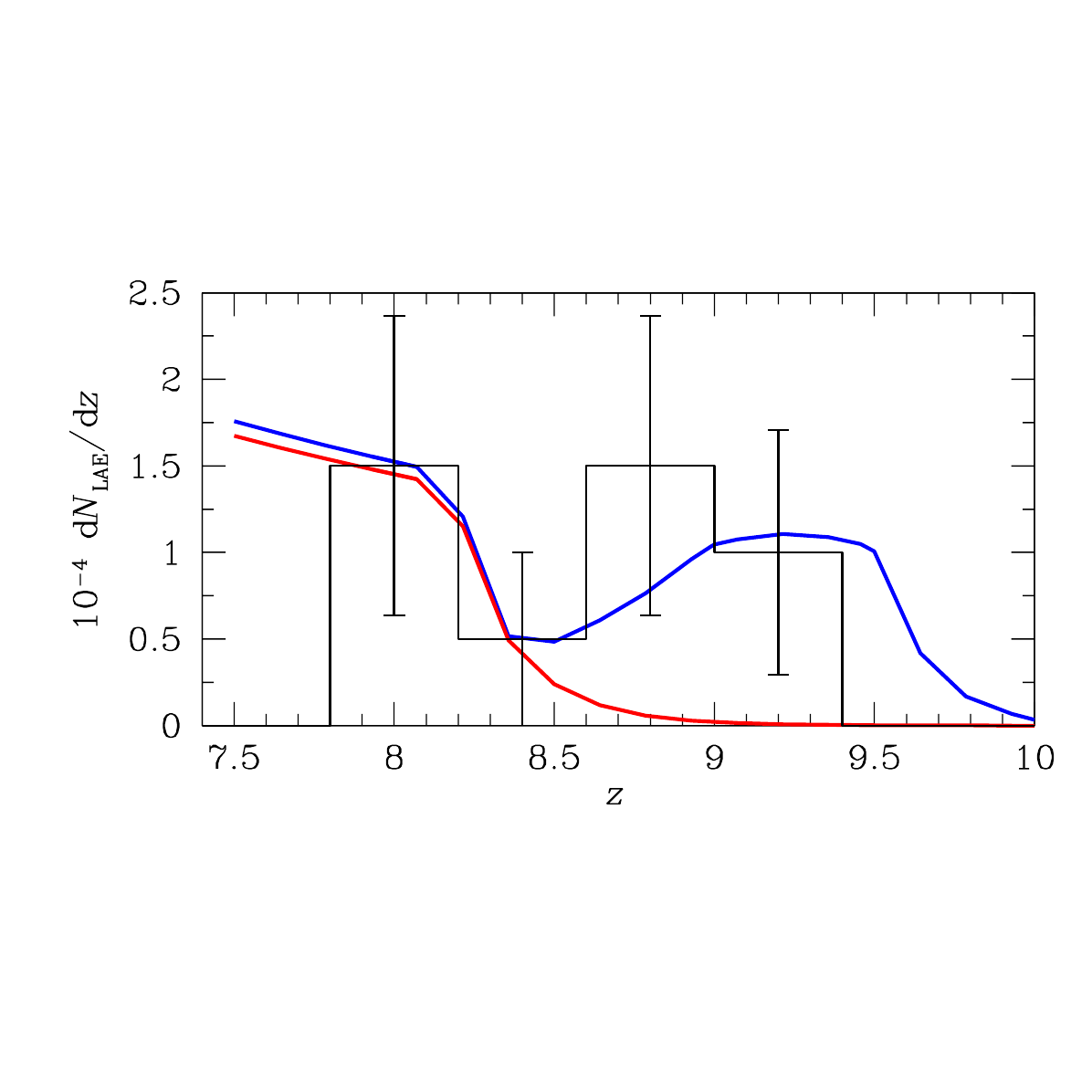}}
\caption{Redshift distribution of LAEs brighter than $L\Lyas=10^{43.5}$ erg s$^{-1}$ per infinitesimal redshift as a function of $z$ predicted in single (red line) and double (blue line) reionization compared to the histogram of real detections The latter has been rescaled so as to match the all-sky predictions in the most populated redshift bin centered at $z=8$. The curve predicted in single reionization, which at that bin below overlaps with that predicted in double reionization, has been slightly shifted downwards for clarity. (A color version of this Figure is available in the online journal.)}
\label{f9}
\end{figure}

In Fig.~\ref{f9} we compare the $z$-distribution of visible bright LAEs predicted in the two reionization scenarios to the observed one. Specifically, we compare the predicted number of visible LAEs brighter than $L\Lyas=10^{43.5}$ erg s$^{-1}$ per infinitesimal redshift, $\der N_{\rm LAE}/\der z=\Phi(>L\Lyas,z)(\der V_{\rm c}/\der z)$, where $\Phi(>L,z)$ is the corresponding cumulative \Lya LF, and $V_{\rm c}$ is the comoving volume per infinitesimal redshift at $z$, to the histogram in redshift bins of $\Delta z=0.4$ width of the 9 observed LAEs with $z>7.75$ listed in section 1. We restrict the histogram to $z>7.75$ because below that redshift there are some LAE pairs \citep{Jea20,Tea20}, which could overestimate the real all-sky distribution (finding the same proportion of pairs in complete all-sky surveys is very unlikely). Strictly speaking, since the observed LAEs do not correspond to systematic surveys in given solid angles but to chance detections, the number of objects observed at each redshift bin cannot be used to infer the number that would be found in all-sky surveys. Nevertheless, since the detections are random, both numbers should be nearly proportional. To better visualize how the histogram compares to the predictions found in the two scenarios, we have rescaled the histogram of observed objects so as to match the predicted all-sky values in the most populated redshift bin centered at $z=8$, where the theoretical predictions coincide.

As can be seen in Fig.~\ref{f9}, the rapid decrease at $z>8.5$ of $\der N_{\rm LAE}/\der z$ in single reionization is clearly in contradiction with observations (as confirmed by the Kolmogorov-Smirnov test). Indeed, the predicted number of bright LAEs per infinitesimal redshift is rapidly decreasing, and becomes negligible by $z\sim 8.5$, while many LAEs are seen up to $z\sim 9.5$. Notice that $z\sim 8.5$ is substantially higher than the value ($z\sim 7.5$) found using the visibility factor. This small discrepancy is certainly due, as mentioned, to the underestimate of $f_{\rm vis}$ through equation (\ref{new}) near zero, but it is also likely due to the fact that the empirical value of $f_{\rm vis}$ fitted at $z\sim 7.3$ is already underestimated because the Schechter fit to the LF used by \citep{Kea18} does not account for bright excess of LAEs with $L\Lyas$ above $L\Lyas^{\min}$ \citep{Meabis15,Zea17,Huea19}. On the contrary, the $z$-distribution predicted in double reionization shows a plateau around $z\sim9.3$ due to the joint effect of the increasing visibility factor and the decreasing galaxy abundance at very high-$z$, which is consistent with the shape of the histogram. In fact, the agreement between predictions and observation is particularly good though it is fortuitous to a great extent given the small number of detections (large error bars), and the heterogeneous selection functions used in the observations. 



A similar result is obtained from the comparison between the predicted and observed UV luminosities of those very high-$z$ LAEs: some of the detected objects appear to have UV luminosities higher than the predicted lower limit in single reionization. Since this test crucially depends on the $f\escG^{\rm ion}$ value adopted in the derivation of the theoretical limit, we plot the theoretical predictions resulting from the favorite value of $f\escG^{\rm ion}=0.054$, and the more conservative one of $f\escG^{\rm ion}$ of 0.1. The UV luminosities of the detected LAEs we plot are those derived by the own authors of the findings. Once again, we only include in the comparison those LAEs detected at $z>7.75$ in order to avoid LAE pairs which could lie in large ionized regions thanks to their combined ionizing luminosity, and be individually fainter than required for isolated objects. In this sense, we must bear in mind that the LAEs with the lowest-redshifts in the sample have more chances to be in groups even if they are apparently isolated, which could explain that they are fainter than expected.

\begin{figure}
\centerline{\includegraphics[scale=0.44]{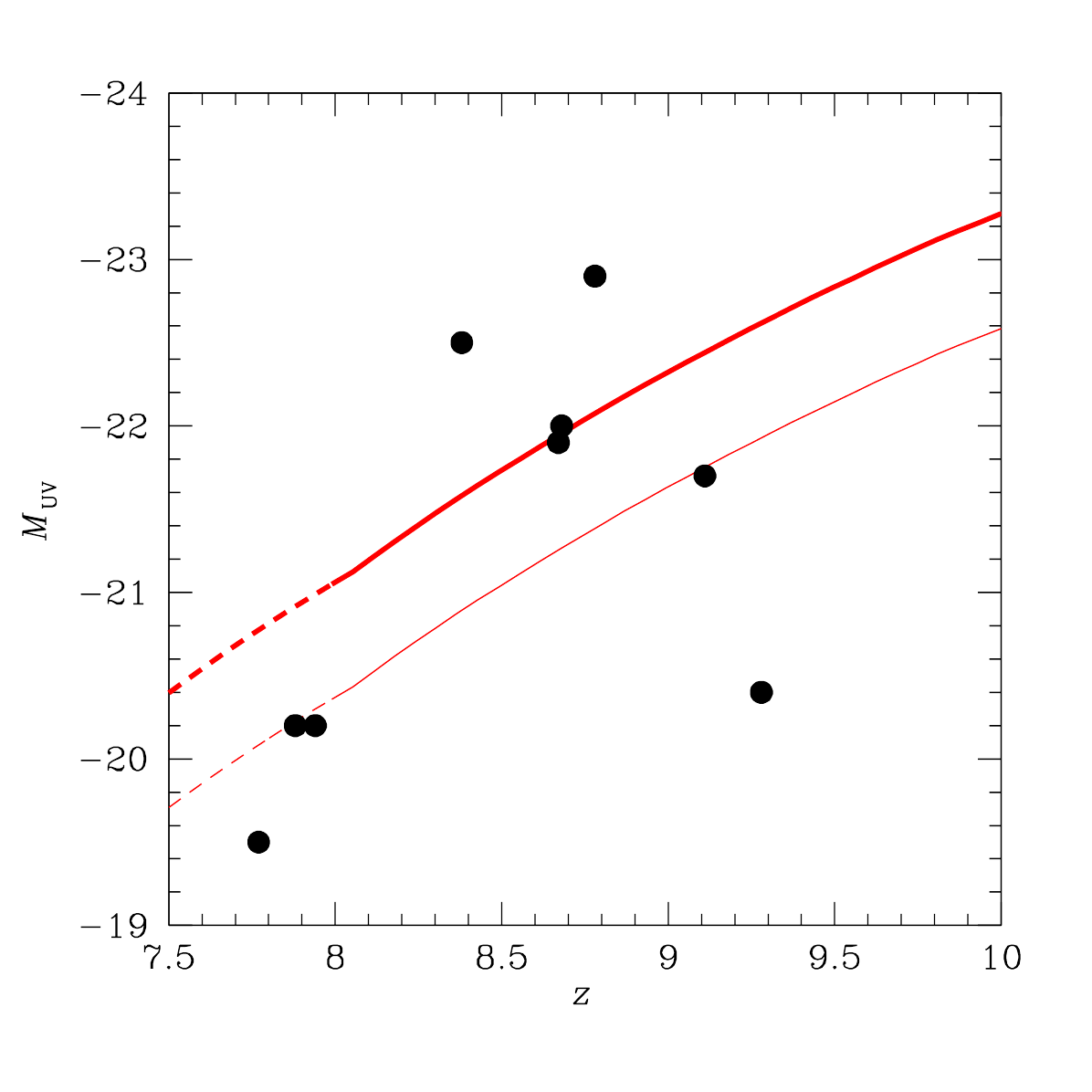}}
\caption{Minimum UV luminosities (actually maximum M$_{\rm UV}$ magnitudes in AB system) of visible LAEs in single reionization as a function of $z$ (red lines) below which there should be no LAE in the single reionization scenario. We show the predictions for the favorite value of $f\escG^{\rm ion}=0.054$ (thick line), and the more conservative value of 0.1 (thin line). Dots mark the UV magnitudes of all LAEs detected at those redshifts, according to the estimates made by the corresponding authors (see the list in section 1). This constraint does not hold in double reionization.}
\label{f10}
\end{figure}

In any event, as shown in Fig.~\ref{f10}, in the favorite case of $f\escG^{\rm ion}=0.054$, only two objects are brighter than $L_{\rm UV}^{\min}(z)$. Two more objects have UV luminosities equal to the lower limit, and 5 objects are fainter, which is very significant. This result is not conclusive, however, because, if we had adopted the more conservative value of $f\escG^{\rm ion}=0.1$, then only two objects would be fainter than $L_{\rm UV}^{\min}(z)$ (and three objects would be at the limit). Moreover, we cannot discard the possibility that $f\escG^{\rm ion}$ is even larger than this, say $\sim 0.2$, in which case only one LAE would be fainter than the lower limit. Of course, the existence of one single object contradicting the lower limit is enough to reject the single reionization scenario, but the derivation of the empirical UV luminosities introduces some uncertainty, so it would be preferable to have more discrepant cases. 

Nevertheless, some aspects of the test reinforce the tentative conclusion that data are inconsistent with single reionization. The UV luminosities of observed LAEs are not increasing with increasing $z$ as expected from the dependency on $z$ of $L_{\rm UV}^{\min}(z)$ (even in the case of a very marked transition from $f\escG^{\rm ion}=0.054$ to $f\escG^{\rm ion}=0.1-0.2$ with increasing $z$). On the contrary, except for the three closest objects with more chances to belong to galaxy overdensities, the rest shows the opposite trend, which is particularly noticeable given that the more distant the objects, the brighter they should tend to be because of the Malmquist bias. In fact, the two most distant LAEs (at $z=9.11$ and $z=9.28$) are those with the lowest (and most discrepant) UV luminosities. As pointed out by \citet{Lea22}, the only reasonable explanation for this surprising fact within the framework of single reionization is that they lie in galaxy overdensities. However, this is quite unlikely at such high-redshifts. Moreover, the fact that these particular LAEs are  lensed objects, which allows one to reach lower luminosities, suggests that the number of discrepant LAEs would likely increase if more lensed objects were observed.

All the previous predictions refer to the abundance of visible LAEs of different luminosities and redshifts, which is the most straightforward LAE property that can be compared to observations. Of course, having determined that abundance, we could also calculate other properties of visible high-$z$ LAEs possible to compare to observation. However, the derivation of any other property would require introducing some specific model calibrated against observations, which would make the final comparison less compelling. More importantly, no other LAE property is expected to be as sensitive to the reionization history as their abundance. Indeed, while this abundance is directly connected to the size of ionized cavities so dependent on the dominant ionizing sources found in each reionization scenario, all intrinsic LAE properties do not. It is true that they may still depend on the reionization history through the distinct abundance of Pop III stars found inside LAEs at any given redshift in both scenarios (this is the case, e.g., of the EW of the \Lya line; \citealt{S03}). However, Pop III stars form prior to normal galaxies, and the abundance of non-exploded objects that are accreted onto galaxies or are lying in the same halos where galaxies develop afterwads is much smaller than that of Pop I and II stars forming in situ, so their possible presence in LAEs should have a very small impact in the intrinsic properties of those galaxies (in fact, they were neglected when calculating the intrinsic \Lya luminosity of LAEs; Sec.~4). Therefore, analyzing other LAE properties would greatly complicate the study for a too small foreseeable gain.

\section{Summary and Conclusions}

In SS+17, we used AMIGA to constrain the reionization history against the observed global (or averaged) properties of luminous objects, the IGM, and the CMB. Only two acceptable solutions were found depending on how top-heavy the unknown Pop III stars IMF was: one with a monotonic reionization process ending at $z=6$, as usually considered, and another non-monotonic one with two full ionization events at $z=6$ and $z=10$. While in the former case ionization is mainly driven by galaxies, in the latter it is by massive Pop III stars. However, when those stars definitely disappear after the first full ionization when the whole IGM is metal-polluted, a recombination phase takes place which lasts until galaxies and AGN take over by $z\sim 7.5$. Since in the Epoch of Reionization the visibility of LAEs greatly depends on the ionization state of the IGM, their observed properties should break the degeneracy between those two reionization scenarios.

To check this we have taken the intrinsic properties of LAEs predicted by AMIGA in those two reionization scenarios, and correct their \Lya luminosities for ISM- and IGM-absorption so as to predict the actual properties of visible very high-$z$ LAEs to be compared to observation. The \Lya LFs so found are very distinct in the two scenarios. In single reionization, they are truncated at the minimum \Lya luminosity of isolated LAEs able to ionize large enough bubbles by themselves, which increases with increasing $z$. This causes visible LAEs to become very rare at $z\ga 8.5$. On the contrary, the LFs in double reionization are not truncated because LAEs lie in ionized cavities that were carved by Pop III stars, uncorrelated with normal galaxies, so that the size of those cavities does not depend on the LAE luminosity. They simply result from multiplying the LFs corrected for ISM-absorption by the visibility factor giving the volume fraction occupied by visible LAEs (i.e. outside neutral regions plus the corresponding \Lya-shadowed zones behind). The comparison of the predicted \Lya LFs with observations at very high-$z$ should thus definitely unravel what is the right reionization scenario. This comparison will soon be possible thanks to new powerful instruments such as the JWST.

At present, one can only compare the predicted $z$-distribution and luminosity of very high-$z$ LAEs with those of a few (9) right objects. One interesting result is that referring to the $z$-distribution of visible very high-$z$ LAEs. In double reionization, the visibility factor increases towards $z=10$, which balances the decreasing galaxy population and the consequent intrinsic LAE abundance. The result is the appearance of a plateau in the abundance of visible bright LAEs between $z\sim 8.75$ and $z\sim 9.5$ consistent with observations. On the contrary, the abundance of visible LAEs in single reionization is predicted to decline with increasing $z$ even faster than the galaxy abundance because the increasing density of the IGM forces LAEs to be increasingly luminous for them to be visible. The result is a predicted $z$-distribution of very high-$z$ LAEs which is in tension with observations. Similarly, the observed UV luminosity of a few of those objects is lower than the minimum UV luminosity predicted in single reionization for isolated visible LAEs. This is particularly the case for the two most distant LAEs, which are lensed objects, and hence, more easy to be seen despite having low luminosities. In this sense, the detection of more lensed LAEs should likely deepen the discrepancy. On the contrary, there is no conflict in double reionization, where visible high-$z$ LAEs are not constrained to have large ionizing luminosities. 

The conclusion is thus that the properties of the very high-$z$ LAEs detected so far slightly favor the double reionization scenario. The incoming data gathered by the new facilities such as the JWST will be crucial to clarify this important issue. 

\begin{acknowledgments}
Funding for this work was provided by the Spanish MCIN/AEI/10.13039/501100011033 under projects CEX2019-000918-M (Unidad de Excelencia `Mar\'\i a de Maeztu' ICCUB), and PID2019-109361GB-100 (co-funded by FEDER funds), and the Catalan DEC grant 2017SGR643. JMMH was funded by Spanish MICIN grants PID2019-107061GB-C61 and MDM-2017-0737 (Unidad de Excelencia `Mar\'\i a de Maeztu' CAB). CC and JG were funded by Spanish MCIN grant AyA2018-rti-096188-b-i00, and JMRE was funded by Spanish MCIN grant AYA2017-84061-P and the Canarian Government under project PROID2021010077. JMRE is also indebted to the Severo Ochoa Program at the IAC.
\end{acknowledgments}


\end{document}